# Millimeter Wave Location-Based Beamforming using Compressive Sensing

Ahmed Abdelreheem, Ehab Mahmoud Mohamed, and Hamada Esmaiel
Electrical Engineering Dept., Aswan University, Aswan, Egypt.
{ahmed.abdelreheem, ehab_mahmoud, and h.esmaiel}@aswu.edu.eg

*Abstract*—This paper develops a location based analog beamforming (BF) technique using compressive sensing (CS) to be feasible for millimeter wave (mmWave) wireless communication systems. The proposed scheme is based on exploiting the benefits of CS and localization to reduce mmWave beamforming (BF) complexity and enhance its performance compared with conventional mmWave analog BF techniques. CS theory is used to exploit the sparse nature of the mmWave propagation channel to estimate both the angle of departures (AoDs) and the angle of arrivals (AoAs) of the mmWave channel, and knowing the node location effectively reduces the number of BF vectors required for constructing the sensing matrix. Hence, a high accurate mmWave BF with a low set-up time can be obtained. Simulation analysis confirms the high effectiveness of the proposed mmWave BF technique compared to the conventional exhaustive search BF and the CS based BF without localization using random measurements.

## I. INTRODUCTION

As the data rate demand grows exponentially, immigration towards high frequency bands seems to be a promising solution. The millimeter wave (mmWave) band provides a large swath of available spectrum that can be used to support high broadband mobile and backhaul traffic, and it can play a major role in the future 5G mobile communication [1]. However, still there are a lot of major challenges towards the use of mmWave technology [2] [3], such as the high propagation path loss and the poor link budget of the mmWave channel and the additional losses due to oxygen absorption and rain [3].

Conventional exhaustive search analog beamforming (BF) techniques can be used to compensate the degradation of the link budget in mmWave without knowing the channel states at the receiver (RX). This can be easily done using directional communication with high antenna gains, where exhaustive search based on a predefined codebook design can be used to examine all available beam directions around a mmWave device and select the TX/RX beam directions maximizing the received power at the RX [4]. Unfortunately, the conventional exhaustive search BF techniques are suffering from low efficiency and high complexity due to the search through all available beam settings. Moreover, a tradeoff between complexity and performance exists depending on the used beamwidth, i.e., as the beamwidth decreases, the BF performance is enhanced while the BF complexity is extremely increased. To overcome this and to relax the complexity of the beam switching process, several techniques have been proposed. Multi-stage codebooks using adaptive beamwidth BF techniques have been presented in [6] – [8]. Although these techniques reduce the required complexity for obtaining the best TX/RX beam directions, they are still exhaustive searching like due to the need to search in all directions around the mmWave device. To enhance the performance of mmWave BF, other researchers, utilizing the sparse nature of the mmWave propagation channel, used compressive sensing (CS) to estimate the channel [9] [10]. Thus, antenna weight vectors can be accurately adjusted in the direction of the estimated angle of departures (AoDs) and angle of arrivals (AoAs) of the estimated channel, which highly increases the BF gain of the antenna array. However, these techniques still face the challenge of how we can efficiently construct the sensing matrix using a small number of beam switching to efficiently estimate the AoDs and the AoAs of the mmWave channel. Wrong positioning of the beams used for taking channel measurements may result in scheme failure due to high errors in the estimated AoDs and AoAs.

A fine beam with a low beam switching complexity are essential demands for a high efficient mmWave BF design. In this way, this paper proposes a localization based mmWave BF using CS to attain this goal. That is, localizing the mmWave devices can extremely relax the problem of selecting the best BF vectors used for channel measurements and constructing the sensing matrix. This is enabled by the fact that mmWave communication is almost a line of sight (LOS) communication. Indoor and outdoor wireless device localization is a hot topic and recently gained a huge attention in wireless communication due to its numerous applications e.g., health care. Various indoor/outdoor localization techniques gather information based on different ways like the long term evaluation (LTE), global positioning system (GPS) or wireless fidelity (WiFi) techniques with an acceptable accuracy [11]. Thanks to localization assistance, this paper relaxes the complexity of CS based mmWave BF, by letting the mmWave TX (RX) devices to probe only in the directions that the RX (TX) devices expected to be localized, respectively. By this way, we highly relax the required complexity for constructing the sensing matrix and the overall complexity of the BF process. At the same time, a high BF gain is expected using the proposed BF scheme.

Via numerical simulations, using the same beamwidth, the proposed scheme outperforms the conventional exhaustive search BF and the CS based BF without localization using random positions of the measurement beams in both BF gain and BF complexity.

The rest of this paper is organized as follows. System model including the optimization problem of mmWave BF and the conventional exhaustive search BF are presented in Sect. II. Section III presents CS based mmWave BF. Section IV will discuss the proposed BF scheme. Section V presents the numerical simulations and analysis, and finally Sect. VI concludes this paper.

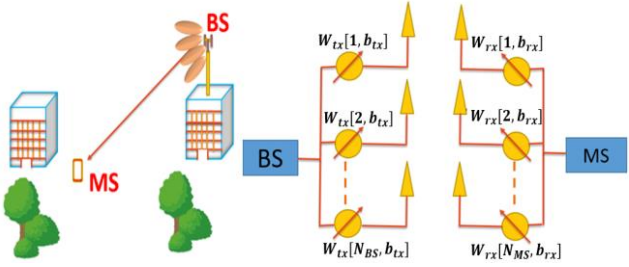

Fig. 1. Block diagram of BS/MS transceiver and mmWave antenna model.

## II. SYSTEM MODEL

This paper considers mmWave transmission between a base station (BS) with an $N_{BS}$ antenna elements and a mobile station (MS) with an $N_{MS}$ antenna elements, and both BS and MS aware of their locations and each other locations, as shown in Fig. 1 (left figure). In mmWave transmissions, the data is up-converted from baseband to radio frequency (RF), and the conventional analog BF is applied to improve the antenna gain as shown in Fig. 1 (right figure). According to the codebook design, the transmitted data are weighted by weight vectors at the TX and RX, as follows [6]:

$$W(n,b) = j^{\text{floor}\left\{\frac{n \times \text{mod}\left(b+\left(\frac{B}{2}\right),B\right)}{\frac{N}{4}}\right\}}, \quad (1)$$

$$n = 0, \ldots, N-1; \ b = 0, \ldots, B-1,$$

where $j = \sqrt{-1}$ and $W(n,b)$ is the antenna weight of antenna element $n$ to accomplish BF in the direction of $b$, where $N$ is the total number of antenna elements and $B$ is the total number of steering beams. In this paper, we assume that both TX and RX antennas have the same number of steering beams. Also, a geometric channel model with $L$ propagation paths is considered between the BS and MS, which can be written as [2]:

$$\mathbf{H} = \frac{1}{\gamma} \sum_{\ell=1}^{L} \mu_\ell \boldsymbol{p}_{MS}(\varphi_\ell) \boldsymbol{p}_{BS}^H(\psi_\ell), \quad (2)$$

where $( )^H$ is Hermitian (conjugate transpose) and $\boldsymbol{p}_{BS}(\psi_\ell)$, $\boldsymbol{p}_{MS}(\varphi_\ell)$ are the array response vectors of the $\ell$ th path at the BS and MS, respectively, which can be written as [2]:

$$\boldsymbol{p}_{BS}(\psi_\ell) = \left[1, e^{j\frac{2\pi}{\lambda}d\sin(\psi_\ell)}, \ldots, e^{j(N_{BS}-1)\frac{2\pi}{\lambda}d\sin(\psi_\ell)}\right]^T, \quad (3)$$

$$\boldsymbol{p}_{MS}(\varphi_\ell) = \left[1, e^{j\frac{2\pi}{\lambda}d\sin(\varphi_\ell)}, e^{j(N_{MS}-1)\frac{2\pi}{\lambda}d\sin(\varphi_\ell)}\right]^T, \quad (4)$$

where $\gamma$ is the average path-loss between BS and MS. $\mu_\ell, \varphi_\ell \in [0,2\pi]$ and $\psi_\ell \in [0,2\pi]$ are the complex gain, the azimuth AoD and the azimuth AoA of the $\ell$ th path, respectively. $d$ and $\lambda$ are the distance between the antenna elements and the signal wavelength respectively. In this paper, 2-D beamforming is considered assuming that all scattering happen in azimuth only. The received signal at MS after applying the BF weight vectors in the directions of $b_{tx}$ and $b_{rx}$ at the BS and MS respectively can be written as:

$$y = \mathbf{W}_{rx}[:,b_{rx}]^H \mathbf{H} \mathbf{W}_{tx}[:,b_{tx}]s + \mathbf{W}_{rx}[:,b_{rx}]^H \boldsymbol{n}, \quad (5)$$

where $y$ and $s$ are the received and transmitted symbols, respectively. $\mathbf{W}_{tx}[:,b_{tx}]$ and $\mathbf{W}_{rx}[:,b_{rx}]$ are the weight vectors of lengths $[N_{BS} \times 1]$ and $[N_{MS} \times 1]$ corresponding to the columns $b_{tx}$ and $b_{rx}$ in the TX cookbook $\mathbf{W}_{tx}$ and the RX codebook $\mathbf{W}_{rx}$, respectively. $\mathbf{H}$ is the mmWave channel matrix of size $N_{MS} \times N_{BS}$ and $\boldsymbol{n}$ is the $[N_{MS} \times 1]$ Gaussian noise vector corrupting the received symbol. Accordingly, the optimization problem of the mmWave BF can be formulated as:

$$(b_{tx}^*, b_{rx}^*) = \arg\max_{1 \le b_{tx}, b_{rx} \le B} |\mathbf{W}_{rx}[:,b_{rx}]^H \mathbf{H} \mathbf{W}_{tx}[:,b_{tx}]|^2, \quad (6)$$

where $b_{tx}^*$ and $b_{rx}^*$ are the TX/RX BF directions corresponding to the columns of the TX/RX codebooks maximizing the channel BF gain.

In the conventional exhaustive searching BF technique, the TX and RX tries all possible $\mathbf{W}_{tx}[:,b_{tx}]$ and $\mathbf{W}_{rx}[:,b_{rx}]$ pairs in the predefined TX and RX codebooks with a total number of beam switching complexity of $B \times B$ assuming that both codebooks have the same total number of steering beams. Due to not estimating the AoDs and AoAs of the mmWave channel, as we increase the number of steering beams, i.e., narrowing the beamwidth, the BF gain of the exhaustive search BF will be enhanced at the expense of highly increasing the beam switching complexity.

## III. CS BASED MMWAVE BF

Thanks to the sparse nature of the mmWave channel, CS can be used to estimate the AoDs and the AoAs of the channel. Towards that, mmWave channel estimation problem should be formulated as a sparse problem as followings.

Let:

$$\mathbf{Y} = \mathbf{W}_{rx}^H \mathbf{H} \mathbf{W}_{tx} \mathbf{S} + \mathbf{W}_{rx}^H \mathbf{n}, \quad (7)$$

where $\mathbf{W}_{tx}$ and $\mathbf{W}_{rx}$ are the BF matrices at the TX and RX of sizes $[N_{BS} \times B]$ and $[N_{MS} \times B]$ respectively. $\mathbf{S}$ is a diagonal matrix carrying the transmitted $B$ symbols, which can be expressed as $\mathbf{S} = \sqrt{P}\mathbf{I}_B$, where $P$ is the TX symbol power, and $\mathbf{I}_B$ is the identity matrix of size $B \times B$. $\mathbf{W}_{rx}^H \mathbf{n}$ is the corresponding noise term. To transfer the mmWave channel estimation problem into a sparse problem, we vectorize the $\mathbf{Y}$ matrix, as follows:

$$\text{vec}(\mathbf{Y}) = \sqrt{P}\,\text{vec}(\mathbf{W}_{rx}^H \mathbf{H} \mathbf{W}_{tx} + \mathbf{W}_{rx}^H \mathbf{n}), \quad (8)$$

$$\mathbf{y}_v = \sqrt{P}\,(\mathbf{W}_{tx}^T \otimes \mathbf{W}_{rx}^H)\,\text{vec}(\mathbf{H}) + \text{vec}(\mathbf{W}_{rx}^H \mathbf{n}), \quad (9)$$

using the channel model given in (2), $\mathbf{y}_v$ can be written as:

$$\mathbf{y}_v = \sqrt{P}\,(\mathbf{W}_{tx}^T \otimes \mathbf{W}_{rx}^H)\,\text{vec}(\mathbf{P}_{BS}^* \circ \mathbf{P}_{MS})\boldsymbol{\mu} + \mathbf{n}_x, \quad (10)$$

where $\mathbf{n}_x = \text{vec}(\mathbf{W}_{rx}^H \mathbf{n})$ and $\mathbf{P}_{BS}^* \circ \mathbf{P}_{MS}$ is the Khatri-Rao product of $\mathbf{P}_{BS}^*$ and $\mathbf{P}_{MS}$ resulting in an $N_{BS}N_{MS} \times L$ matrix, where each column $\ell$ in this matrix represents $(\boldsymbol{p}_{BS}^*(\psi_\ell) \otimes \boldsymbol{p}_{MS}(\varphi_\ell))$ the Kronecker product of BS and MS array responses. To formulate the sparse problem and facilitate the estimation of the AoDs and AoAs of the mmWave channel, AoDs and AoAs values are assumed to be uniformly quantized in a grid of $N$ points, with $N \gg L$, as follwos:

$$\overline{\psi_\ell}, \overline{\varphi_\ell} \in \left\{0, \frac{2\pi}{N}, \ldots, \frac{2\pi(N-1)}{N}\right\}, \ell = 1,2,\ldots,L. \quad (11)$$

where $\overline{\varphi_\ell}$ and $\overline{\psi_\ell}$ are the quantized AoDs and AoAs. Sure, AoDs and AoAs are continuous values, hence a quantization error is expected. Thus the sparse problem can be formulated as:

$$\mathbf{y}_v = \sqrt{P}\,(\mathbf{W}_{tx}^T \otimes \mathbf{W}_{rx}^H)\,\mathbf{T}_D \mathbf{z} + \mathbf{n}_x, \quad (12)$$

where $\mathbf{z}$ is the vector of length $N^2 \times 1$ containing the path gains corresponding to the quantized AoDs and AoAs. $\mathbf{T}_D$ is a dictionary matrix of size $N_{BS}N_{MS} \times N^2$, where each column in this matrix follows $\left(\boldsymbol{p}_{BS}^*(\bar{\psi}_u) \otimes \boldsymbol{p}_{MS}(\bar{\varphi}_v)\right)$,

$$\bar{\psi}_u = \frac{2\pi u}{N}, u = 0,1,...,N-1, \quad (13)$$

$$\bar{\varphi}_v = \frac{2\pi v}{N}, v = 0,1,...,N-1. \quad (14)$$

Estimating a non-zero value in the vector **z** means an estimation of an existing AoD and AoA pair. Besides, this non-zero value is equal to the path gain corresponding to this pair of AoD and AoA. To complete the sparse problem formulation, we assume that **z** has only $L$ non-zeros elements and $L \ll N^2$, now we need to design an efficient sensing matrix $\boldsymbol{\Phi} = \sqrt{P}\,(\mathbf{W}_{tx}^T \otimes \mathbf{W}_{rx}^H)\,\mathbf{T}_D$, to recover the non-zero elements in **z** with a small number of beam switching, which is the core challenge of the CS based mmWave BF. Currently, there are two proposed schemes to tackle this challenge in literature [9] [10]. The first method is the random based measurements, in which the BF vectors used in constructing the $\boldsymbol{\Phi}$ matrix are chosen randomly [10]. However, this method may fail to estimate the AoDs and AoAs in the cases when the BF vectors are not coincident with the quantized angles [10]. The other method is a highly complicated one in which multi-stage beam searching is used to estimate the AoDs and AoAs of the mmWave channel [9]. In this paper, to efficiently construct a high accurate $\boldsymbol{\Phi}$ matrix using a small number of beam switching, localization is utilized.

## IV. PROPOSED LOCATION-BASED MMWAVE BF USING CS

The localization service provides both the BS and the MS with an accurate information about each other locations. This enables the BS (MS) to localize the best measurement BF vectors for constructing $\boldsymbol{\Phi}$ matrix to accurately estimate AoDs (AoAs) of the mmWave channel. Based on localizing the MS (BS), the BS (MS) estimates a range of quantized AoDs (AoAs) $\bar{\psi}_{q_t}$ ($\bar{\varphi}_{g_t}$) expecting to be align with the AoDs (AoAs) of the actual channel. The range of estimated $\bar{\psi}_{q_t}$ and $\bar{\varphi}_{g_t}$ can be expressed as:

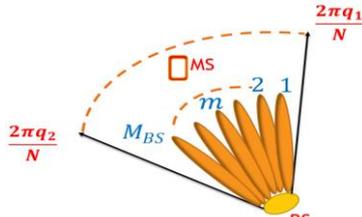

Fig. 2. Design the codebook according to expected range of the AoDs.

$$\bar{\psi}_{q_t} = \{\bar{\psi}_u : q_1 \le u \le q_2, q_1, q_2 \in [0, N-1]\}, \quad (15)$$

$$\bar{\varphi}_{g_t} = \{\bar{\varphi}_v : g_1 \le v \le g_2, g_1, g_2 \in [0, N-1]\}, \quad (16)$$

where $\frac{2\times\pi\times q_1}{N}$ and $\frac{2\times\pi\times q_2}{N}$ are the limits of the estimated AoDs based on the MS location. Similarly, based on the estimated BS location, the limits of the estimated AoAs will be $\frac{2\times\pi\times g_1}{N}$ and $\frac{2\times\pi\times g_2}{N}$ as shown in Fig. 2 of the AoDs case.

Based on localizing $\bar{\psi}_{q_t}$ and $\bar{\varphi}_{g_t}$, the TX/RX precoding matrices can be constructed using multiple BF vectors in the ranges of $[q_1, q_2]$ and $[g_1, g_2]$ for BS and MS, respectively. For BS, assuming $M_{BS}$ different BF vectors with equal beamwidths, each BF vector $m$ in the precoding matrix should satisfy the following, see Fig. 2:

for $m=1$:
$$\mathbf{W}_{tx}[:,m]\boldsymbol{p}_{BS}(\bar{\psi}_u) = \begin{cases} C & if\ u \in \left[q_1, \frac{q_2-q_1}{M_{BS}}\right], \\ 0 & otherwise \end{cases}$$

for $2 \le m \le M_{BS} - 1$:
$$\mathbf{W}_{tx}[:,m]\boldsymbol{p}_{BS}(\bar{\psi}_u)$$
$$= \begin{cases} C & if\ u \in \left[m\frac{q_2-q_1}{M_{BS}}, (m+1)\frac{q_2-q_1}{M_{BS}}\right], \\ 0 & otherwise \end{cases}$$

and for $m=M_{BS}$:
$$\mathbf{W}_{tx}[:,m]\boldsymbol{p}_{BS}(\bar{\psi}_u) = \begin{cases} C & if\ u \in \left[(M_{BS}-1)\frac{q_2-q_1}{M_{BS}}, q_2\right], \\ 0 & otherwise \end{cases} \quad (17)$$

where $C$ is a constant indicate the projection of the BF vector $m$ in the direction of its specified range of quantized AoDs. Same above equations are used for constructing the RX precoding matrix $\mathbf{W}_{rx}$ except that $M_{MS}$, $q_1$ and $q_2$ are replaced by $M_{BS}$, $g_1$ and $g_2$ respectively.

After solving the CS problem using one of the famous CS algorithms like the orthogonal matching pursuit (OMP) [9], which is used in this paper to solve the CS problem, the BF vectors corresponding to the estimated AoDs and AoAs are used for mmWave data transmissions.

## V. SIMULATION ANALYSIS

In this section, we evaluate the performance of the proposed BF scheme using numerical simulations. A downlink data link from the BS to the MS has been considered assuming the mmWave system architecture presented in Fig. 1 (left figure). The BS has $N_{BS} = 8$ antennas, and the MS has $N_{MS} = 8$ antennas using uniform linear array (ULA) antenna. The simulated mmWave channel is a Rician channel with $L = 4$ paths, and the AoDs/AoAs are assumed to be continuous values in the range of $[0, 2\pi]$ with 28 GHz carrier frequency and 100 MHz bandwidth. The maximum ratio of the LOS path to the non-line-of-the sight (NLOS) path is $k = 6$, and a path-loss exponent of $n = 2$ is considered. Assuming a localization service provides the BS and the MS by each other locations; based on which, the BS and the MS can estimate the limits of the AoDs and AoAs, $\frac{2\times\pi\times q_1}{N}$, $\frac{2\times\pi\times q_2}{N}$, and $\frac{2\times\pi\times g_1}{N}$, $\frac{2\times\pi\times g_2}{N}$, respectively. Sure, $q_1, q_2$ and $g_1, g_2$ depend on the probability of localization and the geometry of the BS and MS estimated locations. In the conducted simulations, we draw the BS and the MS randomly in the simulation area, and then we add a random localization error to their exact locations for obtaining their estimated locations, as follows:

$$\hat{X}_{BS} = X_{BS} + \delta X_{BS}, \hat{Y}_{BS} = Y_{BS} + \delta Y_{BS}, \quad (18)$$

$$\hat{X}_{MS} = X_{MS} + \delta X_{MS}, \hat{Y}_{MS} = Y_{MS} + \delta Y_{MS}, \quad (19)$$

where $X_{BS}, Y_{BS}$ and $\hat{X}_{BS}, \hat{Y}_{BS}$ are the exact and estimated locations of the BS in the X and Y directions, respectively, and $\delta X_{BS}, \delta Y_{BS}$ are their corresponding localization errors. $X_{MS}, Y_{MS}, \hat{X}_{MS}, \hat{Y}_{MS}, \delta X_{MS}$ and $\delta Y_{MS}$ are the values for the MS. The values of the localization errors $\delta X_{BS}, \delta Y_{BS}, \delta X_{MS}$ and $\delta Y_{MS}$ are assumed to be uniform random distribution in the range of [0, 5] $m$ with an average value of 2.5 $m$. Based on the estimated locations of the BS and MS and the maximum expected localization error, the ranges $\frac{2\times\pi\times q_1}{N}$, $\frac{2\times\pi\times q_2}{N}$, and $\frac{2\times\pi\times g_1}{N}$, $\frac{2\times\pi\times g_2}{N}$ are adjusted. The beamwidths of the TX/RX BF vectors are also adjusted to be $5^o$. In the conducted simulations, we compare the BF gain against the required number of beam switching of the proposed BF scheme, the exhaustive search BF and the CS only based BF without localization using random directions of the BF vectors. The BF gain can be defined as:

$$\text{Gain} = \max_{B_x} \frac{|\mathbf{W}_{rx}^H \mathbf{H} \mathbf{W}_{tx}|^2}{|\mathbf{H}|^2}, \quad (20)$$

where $B_x$ is the number of examined BF vectors of each different BF method. Figs. 3 and 4 show the cumulative distribution function (CDF) of the BF gains of the proposed scheme, the exhaustive search BF and the CS only BF, respectively. As shown in these figures, for obtaining the same BF gain, the proposed scheme only needs an average value of 14 beam switching, while the exhaustive search BF needs 72 beam switching and the CS only BF needs 50 beam switching. This means that the proposed scheme reduces the complexity of mmWave BF by ratio 80.52 % and 72 % compared to the exhaustive search and CS only BFs while obtaining nearly same BF gain performance. Furthermore, in the case of low $B_x$ with sharp beam CS only BF fails to estimate the AoDs/AoAs correctly, where it is based on random measurements. Hence, in this case, CS only BF has a low performance compare to the exhaustive search BF. To avoid this performance degradation, CS only BF can use wide beams, but unfortunately the range of the covered area will be reduced.

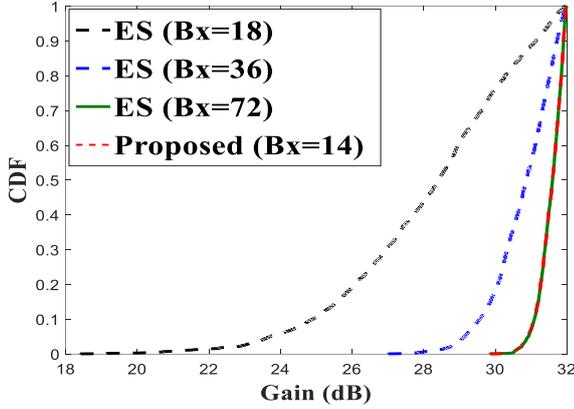

Fig. 3. Performance comparisons between the proposed BF scheme and the exhaustive search (ES) BF.

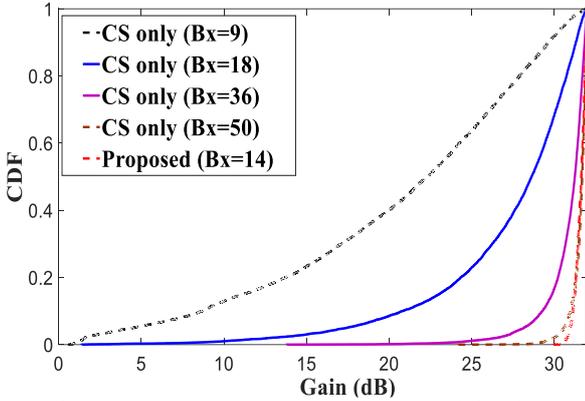

Fig. 4. Performance comparisons between the proposed BF scheme and the CS only BF.

## VI. CONCLUSION

This paper developed a localization based mmWave BF technique using CS for mmWave wireless communication systems. Based on localization, the AoDs/AoAs searching ranges will be reduced and sharp beams can be utilized, which contribute in highly reducing the number of BF vectors needed to accurately estimate the actual AoDs/AoAs of the channel. Thanks to the proposed localization based mmWave BF, the proposed scheme succeeded to reduce the required beam switching complexity by 80.52 % and 72 % for obtaining nearly same BF gain like the exhaustive search BF and the CS only BF without localization, respectively. Further investigations on optimizing the beamwidth of the BF vectors in addition to the searching angles of the AoDs/AoAs based on localization will be the motivation of our future work.

ACKNOWLEDGMENT

This work is partially supported by National Telecom Regulatory Authority (NTRA) Egypt under project title "LTE/WiFi/WiGig internetworking for future 5G cellular networks".